# Time-dependent density-matrix functional theory for trion excitations: application to monolayer MoS$_2$

Alfredo Ramirez-Torres, Volodymyr Turkowski, and Talat S. Rahman[*]

Department of Physics, University of Central Florida, Orlando, FL, 32816

*Corresponding author, e-mail address: talat@physics.ucf.edu

We study possible optically excited bound states in monolayer MoS$_2$: excitons and trions. For this purpose we formulate and apply a generalized time-dependent density-matrix functional approach for bound states of multiple excitations. The approach was used in the cases of three different types of the exchange-correlation (XC) kernel: 1) two local kernels – a phenomenological contact and the adiabatic local-density approximation (ALDA) (X and XC); 2) gradient-corrected X kernels: GEA, PW91 and PBE; and 3) two long-range (LR) kernels: a phenomenological (Coulomb) and Slater kernels. In the case of exciton, we find that LDA and its gradient-corrected kernels lead to too weak binding energy comparing to the experimental data, while the LR kernels are capable to reproduce the experimental results. Similarly, in the LR case (as well as in the case of local kernel), one can obtain the experimental value of the trion binding energy by taking into account the screening effects. Our results suggest that similar to the excitons, the LR structure of the XC kernel is necessary to describe the trion bound states. Our calculations for the first time confirm theoretically with time-dependent density-functional theory approach that in agreement with experimental data the exciton and trion binding energies are of order of hundreds (excitons) and tenth (trions) meVs, which can be used in different technological applications at the room temperature regime. The approach can be straightforwardly extended on the case of bound states and nonequilibrium response of systems with larger number of bound electrons and holes, including biexcitons.

PACS numbers:78.67.-n,73.21.b,71.10.-w, 71.15.Mb

## I. Introduction

Studies of the physical properties of MoS$_2$ monolayer is a hot topic in both experimental and theoretical areas (see, e.g., Refs. [1-12]). These studies is a part of the exploration of new types of two-dimensional materials which potentially extend the fascinating properties of graphene. Being discovered rather recently,[1] this transition-metal chalcogenide system has already recommended itself as a very promising candidate for new nanotechnological applications. In particular, contrary to the bulk case, monolayer $MoS_2$ is a direct gap semiconductor (with the optical gap of 1.8eV at K-points) with a very high quantum efficiency for the luminescence.[1,2] The system also demonstrates a high electron mobility, room-temperature current on/off ratio and ultralow standby power dissipation, with potential to be used in field-effect transistors.[3] It was demonstrated that one can achieve complete dynamic (longer than 1ns) valley polarization in the monolayer $MoS_2$ by optical pumping it with a circularly polarized light.[4,5] Control of the polarization in two direct gap energy valleys (at K and K' points) demonstrates possibility of the system to be used in valley-based electronic and optoelectronic devices. Due to the reasons mentioned above, the optical properties of the system are of a special interest. In particular, possibility of excitonic and higher order excited bound states needs an accurate study. Experimental data suggests that there are strong indications of the excitonic effects in this system with very large (~1eV) binding energy.[1,2,6] Recently, another exciting property of the $MoS_2$ monolayer was discovered, namely trion bound states with the binding energy approximately 20meV.[7] Due to a large binding energy, both excitonic and trionic effects may have applications at room temperatures. While theoretical studies of the excitonic effects in monolayer $MoS_2$ were performed by some groups (see, e.g., Refs. 8,9,12 were the phenomenological Wannier equation and the GW/Bethe-Salpeter equation (GW/BSE) approaches were used), the theoretical studies of the trion effects in this system are very limited. In work [12], a trial wave function approach to calculate both the exciton and trion binding energies. The authors demonstrated that one can obtain the binding energies for both quasi-particles in a reasonable agreement with experimental data (though ~30%-over-estimated for trions). On the other hand, such an approach is limited mostly to study the bound states, while for study of the excitation dynamics TDDFT is a much better candidate (see below).

In general, accurate description of the bound states and their dynamics in semiconductors, is a rather complicated task. Exciton and trion are among the most important bound states. Formally, exciton is defined as a coupled electron-hole pair, while trion is a bound state of an exciton and an electron, so the trion can be regarded as a charged exciton (Figure 1). While in the case of bulk systems the trion binding energy is typically much smaller comparing to the exciton one, in the case of constrained geometry this energy can be pronounceable, which may lead to important effects. The most important example is the case of quantum wells, where trion excitations affect the optical,[10,11] transport[13] and diffusion properties of the system.[14]

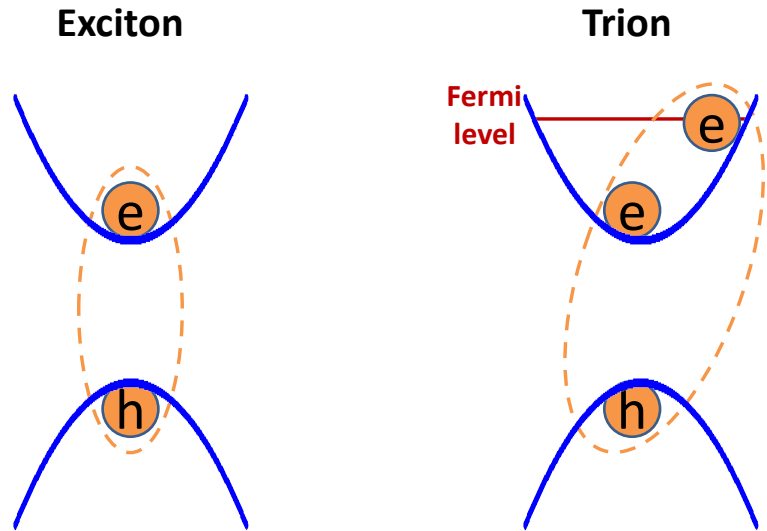

Figure 1. Schematic representation of the exciton (left) and trion (right) quasi-particles in the case of a two-band model. In the trion case, it is assumed that the exciton (electron-hole pair with energies near the bands edges, i.e. with total zero momentum) is coupled to another electron at the Fermi level, the case considered in the paper.

In the case of excitons, the standard time-dependent Hartree-Fock approximation leads to strongly over-bound states, which stimulated development of more subtle many-body methods that take into account correctly screening and other many-particle effects, most known of which is based on the GW/BSE approach.[15] Unfortunately, this approach becomes very complicated in the case of multiple bound states (trions, biexctions, etc.) and when one needs to study strongly nonequilibrium processes (for example, the ultrafast response), that demand to deal with Green's functions that depend on many time arguments. From this point of view, TDDFT[16] is much better candidate. Being a theory of one function – a space- and time-dependent electron charge density, it allows one to get a simple accurate numerical solution of the system response, provided that the corresponding XC potential describing the interaction effects is known.[17] Some progress in incorporating the excitonic effects into TDDFT has already been made. In particular, in Refs. 15,18,19 the corresponding XC kernel was constructed by using a many-body Green's function approach, while in Refs. 20,21 the exact-exchange approximation was used for this purpose. Despite a good agreement with the experiments, these methods are not less complicated than the many-body ones. Recently, we have proposed a technically much simpler and physically transparent TDDFT approach to study excitonic and biexcitonic effects.[22-24] The approach is based on density-matrix representation of the electron wave-function, and the corresponding equations for the excitonic and biexcitonic transition matrix elements (generalized TDDFT Bloch equations) allow one to calculate the excitonic and biexcitonic binding energies. In particular, the exciton equation can be regarded as the TDDFT version of the many-body exciton Wannier equation. We have demonstrated that in principle one can obtain a rather good agreement with experimental data by choosing proper XC kernel (for example, local or LR phenomenological kernels).

To the best of our knowledge, theoretical studies of the trions were performed exclusively in terms of effective many-body models (see, e.g., Refs. 12, 25-28). Similar to the case of excitons, while many-body approaches can lead to a qualitative and sometimes a quantitative description of the trion effects, it is desirable to have a simple and transparent method to describe quantitatively the trion properties of real systems, including the nonequilibrium case. In this work we generalize our TDDFT approach on the case of trions, and show that the corresponding theory gives rather good agreement with experimental values for the excitonic and trion binding energies for monolayer $MoS_2$ in the case of some XC potentials.

# II. TDDFT approach for trions

To derive the TDDFT equation for the trion binding energy, which we define as the energy necessary to decouple one electron from the coupled electron-hole pair (exciton), we begin with a summary of our density-matrix TDDFT approach for the exciton and biexciton bound states (more details can be found in Refs. 22-24).

a) Excitons. In the case of excitons, one can proceed from the Kohn-Sham equation

$$i\frac{\partial \Psi_k^v(\mathbf{r}, t)}{\partial t} = H(\mathbf{r}, t)\Psi_k^v(\mathbf{r}, t), \tag{1}$$

where the system Hamiltonian

$$H(\mathbf{r}, t) = -\frac{\nabla^2}{2m} + V_H[n](\mathbf{r}, t) + V_{XC}[n](\mathbf{r}, t) + e\mathbf{r}\mathbf{E}(t) \tag{2}$$

includes the kinetic (first), Hartree (second) and XC (third) potential terms, as well as the external homogeneous electric field (the last term). Equation (1) is solved self-consistently with the number equation:

$$n(\mathbf{r}, t) = \sum_{l, |\mathbf{k}| < k_F} \left|\Psi_k^l(\mathbf{r}, t)\right|^2. \tag{3}$$

To solve Eqs.(1), (2) it is convenient to use the density-matrix formalism[22] in which the wave function is expanded in terms of the basis (e.g., Bloch) static wave functions $\psi_k^l(r)$:

$$\Psi_k^v(\mathbf{r}, t) = \sum_l c_k^l(t)\psi_k^l(r),$$ (4)

were **k** is the momentum and l is the band index. The time-dependent coefficients $c_k^l(t)$ completely describe the system dynamics. They can be found from the following equation:

$$i\frac{\partial c_k^m}{\partial t} = \sum_k H_{kk}^{ml}c_k^l,$$ (5)

were

$$H_{kq}^{lm}(t) = \int \psi_k^{l*}(r)H[n](r,t)\psi_q^m(r)dr.$$ (6)

However, to study the system response it is more convenient to consider the bilinear combination of c-coefficients, the density matrix:

$$\rho_{kq}^{lm}(t) = c_k^l(t)c_q^{m*}(t).$$ (7)

Its diagonal elements describe the level occupancies, while the non-diagonal – the electron transitions, including the excitonic effects. The matrix elements satisfy the Liouville equation:

$$i\frac{\partial \rho_{kq}^{lm}(t)}{\partial t} = [H(t), \rho(t)]_{kq}^{lm}.$$ (8)

In the case of two (valence v and conduction c) bands, one can derive the exciton TDDFT equation for the non-diagonal element $\rho_{kq}^{cv}(t)$ by using Eqs.(2), (3),(7) and (8).[19] Expansion of the charge density fluctuations in (8) in terms of the density matrix elements (6) (by using Eq. (3)) leads to the TDDFT Wannier equation:[23]

$$\left[(\varepsilon_{k+q}^c - \varepsilon_k^v)\delta_{kk'} + F_{kkk'k'}^{cvvc}\right]\rho_{n,k'+\alpha q}^{cv}(\omega) = E_{n,q}\,\rho_{n,k+\alpha q}^{cv},$$ (9)

where **q** is the exciton momentum, α is the reduced hole mass, and n is the bound state number. The effective electron-hole interaction is described by the last matrix elements defined as:

$$F_{kqk'q'}^{abcd}(\omega) = \int dr_1 dr_2 \psi_k^{a*}(r_1)\psi_q^b(r_1)f_{XC}(r_1, r_2, \omega)\psi_{k'}^{c*}(r_2)\psi_{q'}^d(r_2).$$ (10)

In the case **q**=0 one can obtain the excitonic binding energies from Eq.(9).

b) Biexcitons. Similar, one can consider two-electron TDDFT problem in order to derive the equation for biexcitonic states.[24] In the TDDFT language, this is a problem of two excited electrons in the field of two holes. The corresponding equation is

$$i\frac{\partial \Psi_{k_1 k_2}^{vv}(r_1, r_2, t)}{\partial t} = \left[ H(r_1, t) + H(r_2, t) + \frac{1}{|r_1 - r_2|} \right] \Psi_{k_1 k_2}^{vv}(r_1, r_2, t), \tag{11}$$

where the single-electron Hamiltonian is defined in Eq.(2), while the last term in brackets on the right hand side describes the electron-electron repulsion. The two-particle wave function can be expanded in terms of two single-electron functions:

$$\Psi_{k_1 k_2}^{vv}(r_1, r_2, t) = \sum_{l,m} B_{k_1 k_2}^{lm}(t)\psi_{k_1}^l(r_1)\,\psi_{k_2}^m(r_2), \tag{12}$$

where the two-electron matrix elements satisfy:

$$i\frac{\partial B_{k_1 k_2}^{cd}}{\partial t} = \sum_{a,p}\left[ H_{k_1 p}^{ca} B_{p k_2}^{ad} + H_{k_2 p}^{da} B_{k_1 p}^{ca} \right] + \sum_{a,b,p_1,p_2} w_{k_1 k_2 p_1 p_2}^{cdab} B_{p_1 p_2}^{ab}, \tag{13}$$

with $H_{kp}^{ca}$ defined in Eq. (7) and

$$w_{k_1 k_2 p_1 p_2}^{cdab} = \int dr_1 dr_2\, \psi_{k_1}^{c*}(r_1)\psi_{k_2}^{d*}(r_2)\frac{1}{|r_1 - r_2|}\psi_{p_1}^a(r_1)\psi_{p_2}^b(r_2) \tag{14}$$

is the matrix element that corresponds to the electron-electron repulsion. Similar to the excitonic case, in order to get the biexciton eigen-energies one can consider a linearized form of the corresponding Equation (13). Indeed, if the lowest eigen-energy of this equation is smaller than the sum of two exciton energies obtained from Eq. (9) this means that two excitons form a bound state.

c) Trions. In a similar way one can study the problem of a trion - two excited electrons described by the field $B_{k_1 k_2}^{ab}(t)$ in presence of the hole $c_q^{c*}(t)$. The corresponding matrix element

$$t_{k_1 k_2 q}^{abc}(t) = B_{k_1 k_2}^{ab}(t)c_q^{c*}(t) \tag{15}$$

defines the time-dependence of the three-particle wave function:

$$\Psi_{k_1 k_2 q}^v(r_1, r_2, r_3, t) = \sum_{l,m} t_{k_1 k_2 q}^{lmn}(t)\psi_{k_1}^l(r_1)\,\psi_{k_2}^m(r_2)\psi_q^{n*}(r_3) \tag{16}$$

(the trion excitation corresponds to the upper index $lmn=ccv$). Using Eqs.(5) and (13), one can obtain the following equation for the three-particle density matrix:

$$i\frac{\partial t^{abc}_{k_1 k_2 q}}{\partial t} = \sum_{f,p}\left[H^{af}_{k_1 p}t^{fbc}_{pk_2 q} + H^{bf}_{k_2 p}\,t^{afc}_{k_1 pq} - H^{fc}_{pq}t^{abf}_{k_1 k_2 p}\right] + \sum_{f,m,p_1,p_2}w^{abfm}_{k_1 k_2 p_1 p_2}t^{fmc}_{p_1 p_2 q}, \tag{17}$$

where the H- and w-matrix elements are defined in Eqs. (6) and (14), correspondingly. Linearization of this equation gives the equation for the trion eigen-energies:

$$i\frac{\partial t^{ccv}_{k_1 k_2 q}}{\partial t} = \left(\varepsilon^c_{k_1} + \varepsilon^c_{k_2} - \varepsilon^v_q\right)t^{ccv}_{k_1 k_2 q}$$
$$+ \sum_{p_1,p_2}\left[F^{cvvc}_{k_1 q p_2 p_1}t^{ccv}_{p_1 k_2 p_2} + F^{cvvc}_{k_2 q p_2 p_1}t^{ccv}_{k_1 p_1 p_2} + w^{cccc}_{k_1 k_2 p_1 p_2}t^{ccv}_{p_1 p_2 q}\right], \tag{18}$$

were $\varepsilon^c_k$ and $\varepsilon^v_k$ are the free electron and free hole spectra, and F and w potentials describe the TDDFT electron-hole and electron-electron scattering (in particular, the first F-term describes the scattering of the first electron with momentum $k_1$ on the hole with momentum q, and similarly the second F-term describes the scattering of the second electron with momentum $k_2$ on the same hole with momentum q).

Eqs. (9), (10), (14) and (18) suggest the following wayof generalization of the corresponding eigen-energy equations on the case of larger number of bound electrons and holes. One can study the possibility of bound states in such systems by using a many-particle Schroedinger equation of type (18), where the electrons and holes attract each other with the potentials, or rather scattering matrix elements F (Eq.(10)), while the electron-electron and hole-hole repulsion potentials are defined by matrix elements w, Eq.(14). Namely, each pair of electrons interact through the TDDFT scattering potential:

$$w^{cccc}_{kq;k'q'} = \int d\boldsymbol{r}_1 d\boldsymbol{r}_2\,\psi^{c*}_k(\boldsymbol{r}_1)\psi^{c*}_q(\boldsymbol{r}_2)\frac{1}{|\boldsymbol{r}_1 - \boldsymbol{r}_2|}\psi^c_{k'}(\boldsymbol{r}_1)\psi^c_{q'}(\boldsymbol{r}_2), \tag{19}$$

which describes the scattering of two electrons with momenta k, q to the states with momenta $k'$ and $q'$. Similarly, one can describe the corresponding hole-hole scattering by changing all band indices from "c" to "v" in the last equation. The electron-hole attraction is described by the scattering potential

$$F^{cvvc}_{kq;k'q'} = \int d\boldsymbol{r}_1 d\boldsymbol{r}_2\,\psi^{c*}_k(\boldsymbol{r}_1)\psi^v_q(\boldsymbol{r}_2)\frac{1}{|\boldsymbol{r}_1 - \boldsymbol{r}_2|}\psi^c_{k'}(\boldsymbol{r}_1)\psi^{v*}_{q'}(\boldsymbol{r}_2), \tag{20}$$

which similarly describes the scattering of the electron-hole pair from the state with momenta k and q to the state with momenta $k'$ and $q'$. For example, in the case of biexciton (two electrons with momenta $k_1$ and $k_2$ and two holes with momenta $q_1$ and $q_2$) the corresponding equation for the "wave function" $B^{ccvv}_{k_1 k_2 q_1 q_2}$ has the following form:

$$i\frac{\partial B_{k_1 k_2 q_1 q_2}^{ccvv}}{\partial t} = \left(\varepsilon_{k_1}^c + \varepsilon_{k_2}^c - \varepsilon_{q_1}^v - \varepsilon_{q_2}^v\right) B_{k_1 k_2 q_1 q_2}^{ccvv}$$
$$+ \sum_{k,q}\left[F_{k_1 q_1 qk}^{cvvc} B_{kk_2 qq_2}^{ccvv} + F_{k_1 q_1 qk}^{cvvc} B_{kk_2 q_1 q}^{ccvv} + F_{k_2 q_1 qk}^{cvvc} B_{k_1 kqq_2}^{ccvv} + F_{k_2 q_2 qk}^{cvvc} B_{k_1 kq_1 q}^{ccvv}\right]$$
$$+ \sum_{p_1,p_2}\left[w_{k_1 k_2 p_1 p_2}^{cccc} B_{p_1 p_2 q_1 q_2}^{ccvv} + w_{q_1 q_2 p_1 p_2}^{vvvv} B_{k_1 k_2 p_1 p_2}^{ccvv}\right]. \tag{21}$$

This equation has to be compared with the standard Schrödinger equation for two electrons and two holes:

$$i\frac{\partial \Psi_{k_1 k_2 q_1 q_2}}{\partial t} = \left(\varepsilon_{k_1}^c + \varepsilon_{k_2}^c - \varepsilon_{q_1}^v - \varepsilon_{q_2}^v\right) \Psi_{k_1 k_2 q_1 q_2}$$
$$- \sum_{k}\frac{1}{k^2}\left[\Psi_{kk_2-kq_2} + \Psi_{kk_2 q_1-k} + \Psi_{k_1 k-kq_2} + \Psi_{k_1 kq_1-k}\right]$$
$$+ \sum_{p_1,p_2}\frac{1}{k^2}\left[\Psi_{k-kq_1 q_2} + \Psi_{k_1 k_2 k-k}\right]. \tag{22}$$

While the solution of this equation might be not less complicated than the solution of the corresponding many-body equation, the main advantage of the TDDFT approach is inclusion of the many-body effects through the two-particle attraction defined by in principle exact XCkernel. This property is especially important in the strongly non-equilibrium regime with multiple excitations, where non-linear effects must be taken into account.

# III. Excitons and trions in monolayer MoS$_2$

### III.a The method

To study exciton and trion binding energies, we use the Quantum-ESPRESSO package[29] to generate the Kohn-Sham eigenfunctions and eigenenergies, and the BEE (Binding Energies of Excitons) code developed in our group to solve the exciton and trion eigenenergy equations (9) and (18). In DFT stage, the exchange and correlation effects are included by using to LDA by using the Perdew and Zunger parametrization.[30] In these calculations, we also use norm conserving pseudo-potentials[31] and the cutoff energy of 60 Ry. For the reciprocal space calculations, we use the Monkhorst-Pack scheme.[32] A supercell with (1x1) periodicity is used to model a monolayer MoS$_2$, see figure 1. We employ the in-plane lattice parameter a=3.169 obtained from the bulk optimization with a 15x15x15 **k**-point grid, while the distance between the periodic images along the c-axis is 15 Å. In the self-consistent calculations of the relaxed monolayer MoS$_2$ structure, we used 36x36x1 **k**-point grid (217 independent **k**-points in the first Brillouin zone).

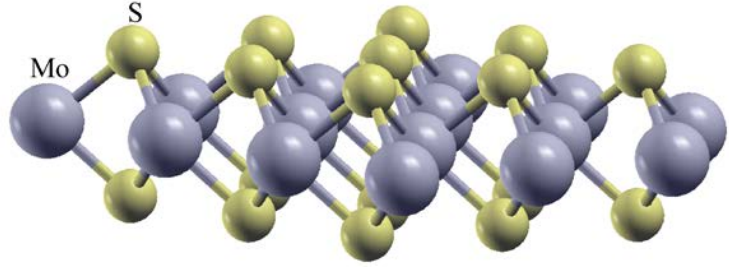

Figure 1. Structure of monolayer $MoS_2$ (grey – Mo atoms, yellow – S atoms).

The results of the band structure calculations are presented in Fig.2.

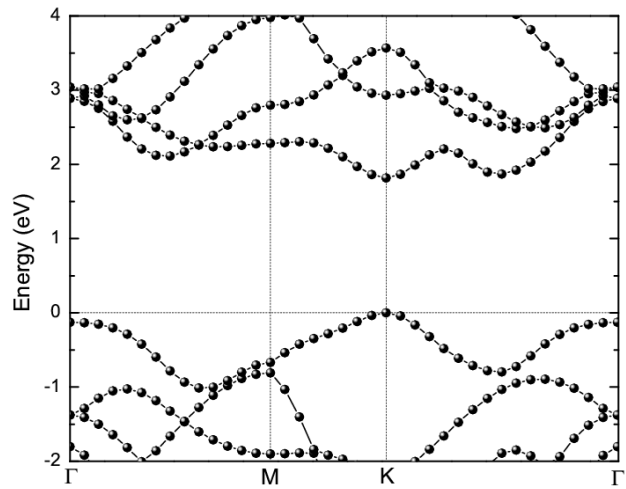

Figure 2. Band structure of monolayer $MoS_2$ calculated with LDA approximation.

As it follows from this Figure, despite the standard underestimation of the bandgap by LDA, our result (1.8 eV) is in a reasonable agreement with experimental estimations.[1] On the other hand, the absolute value of the gap is not essential in this approach, since the binding energy is calculated with respect to the conduction band edge.

To calculate the exciton and trion binding energies we solve equations (8) and (18) in the case of eight kernels:

- Three local: phenomenological contact interaction $f_{XC}^{local}(\boldsymbol{r}, \boldsymbol{r}') = -4\pi A\delta(\boldsymbol{r} - \boldsymbol{r}')$ (A is a parameter describing the strength of the TDDFT local electron-hole attraction) and LDA (both X and XC). Taking into account the correlation part in the case of LDA, we study how the correlations affect the binding energies. One may regard the local kernel as 2D LDA(X) at A=1 (though, obviously, $MoS_2$ is far from being a 2D system, study of this case may help to get an insight how spatial constrain affects the bound state energies).

- Three gradient-corrected kernels: GEA, and two GGA X (PW91 and PBE). These kernels take into account the effects of possible strong spatial variation of the electronic charge, and hence the spatial-dependence of the local electron-hole interaction (which might be important in systems with strongly decreasing at boundaries charge, as MoS2).

- Two LR kernels: phenomenological $f_{XC}(\boldsymbol{r}, \boldsymbol{r'}) = -\frac{1}{\varepsilon} \frac{1}{|r-r'|}$ and the Slater kernel (optimized effective potential (OEP) case) with physically correct electron-hole interaction, which includes a Coulomb singularity (for details on the potentials, see, e.g., Ref. [17] and references therein).

### III.b. Excitons

In the case of excitons, we have found that the contact kernel can reproduce the experimentally estimated energy 1eV[8] at A=0.395 (we don't include the spin-orbital band splitting into account which results in two corresponding exciton peaks, the generalization of the results on this case is trivial). The binding energy is very sensitive to the value of A (Table I). On the other hand the LDA approximation gives very small binding energies in both X and XC cases: 2.05meV and 2.0meV, correspondingly (Table II). Comparing this result to the contact kernel one, and taking into account the fact that required A, equal 0.395, is of order 1 (2D LDA), one can suggest that indeed the spatial constraint in one direction may be important for the excitons in this system. On the other hand, very small decrease of the LDA binding energy with inclusion of the correlations suggests that the correlation effects are not very important in this system (though the situation may change dramatically when, e.g., one dopes the system with transition metal atoms). We also found that charge-gradient correction does not improve the situation significantly. The GEA and PW91 binding energies are even lower than LDA ones. Though PBE gives much larger energy than LDA, it is still much lower than the experimental value.

Table I. Exciton binding energy (in meV) in the case of local approximation for XC at different values of coupling A.

| A | 1 | 0.5 | 0.395 | 0.238 | 0.213 | 0.1 | exp |
|---|---|-----|-------|-------|-------|-----|-----|
| $E_X$ | 3,863 | 1,494 | 1,000 | 300 | 200 | 1 | 1,000 |

Table II. Exciton binding energy (in meV) in the case of different kernels. In the LDA case both X and XC results are presented, while in the GEA, PW91 and PBE cases only the X result is shown. In the LR case, $\varepsilon = 1$.

| Kernel | LDA (X) | LDA (XC) | GEA | PW91 | PBE | LR | Slater | exp |
|--------|---------|----------|-----|------|-----|-----|--------|-----|
| $E_X$ | 2.05 | 2.0 | 0.87 | 1.96 | 10.46 | 90.54 | 1,093 | 1,000 |

On the other hand, the result changes dramatically when one takes into account the Coulomb nature of the interaction (a kernel with $\frac{1}{q^2}$ singularity (see, e.g., Refs. [20-22])). While the unscreened ($\varepsilon = 1.0$) phenomenological LR kernel gives somewhat underestimated value of the binding energy (90.54meV), the Slater result 1093meV is in rather good agreement with the experiment. Despite the fact that in the first case the results are very sensitive to the value of the screening parameter $\varepsilon$ and one can obtain the experimental energy by "artificially" lowering the screening parameter by two times comparing to the vacuum (unscreened) value, it seems problematic to get accurate experimental value if one uses experimentally motivated values of the screening parameter: the parallel and perpendicular components of the dielectric constant: $\varepsilon_\parallel = 2.8$ and $\varepsilon_\perp = 4.2$ (Ref. [10]), and their average: $\bar{\varepsilon} = \sqrt{(2\varepsilon_\parallel^2 + \varepsilon_\perp^2)/3} \approx 3.3$, which corresponds also to the dielectric constant of bulk $MoS_2$ (Ref. [33]). $E_X$ is extremely sensitive to the value of screening at $0.5 < \varepsilon < 3 - 4$ (Table III). It suggest that this potential will result accurate description of the exciton effects when used as a part of a hybrid potential, for example with one of GGAs

Table III. Exciton binding energy (in meV) in the case of LR approximation and different values of $\varepsilon$.

| $\varepsilon$ | 0.449 | 0.752 | 0.844 | 1 | 2.8 | 4.2 | 3.3 | exp |
|---|---|---|---|---|---|---|---|---|
| $E_X$ | 1,000.00 | 300 | 200 | 90.54 | 0.62 | 0.39 | 0.51 | 1,000.00 |

### III.c. Trions

Equation (18) for the trion energy is rather complicated to be solved exactly, therefore we use an approximation, similar to the many-body case. Namely, it is convenient to reduce the problem to a problem of electron with momentum $\boldsymbol{k}_1$ in presence of an exciton, made of the remaining electron and hole (momenta $\boldsymbol{k}_2$ and $\mathbf{q}$). In this case, using Eq.(9) for the exciton function, one can transform Eq.(18) to

$$i\frac{\partial t^{ccv}_{k_1 k_2 q}}{\partial t} = \left(\varepsilon^c_{k_1} + E_{X\,k_2,q}\right) t^{ccv}_{k_1 k_2 q} + \sum_{p_1,p_2}\left[F^{cvvc}_{k_1 q p_2 p_1} t^{ccv}_{p_1 k_2 p_2} + w^{cccc}_{k_1 k_2 p_1 p_2} t^{ccv}_{p_1 p_2 q}\right]. \tag{23}$$

Next, we assume that the excitonic electron and hole momenta are fixed, $\boldsymbol{k}_2 = \boldsymbol{q}$, i.e. we consider the exciton with fixed center-of-mass. In this case, the trion equation reduces to

$$i\frac{\partial t_{k_1qq}^{ccv}}{\partial t} = (\varepsilon_{k_1}^c + E_{X\,q,q})t_{k_1qq}^{ccv} + \sum_{p_1,p_2}\left[F_{k_1qqp_1}^{cvvc}t_{p_1qq}^{ccv} + w_{k_1qp_1q}^{cccc}t_{p_1qq}^{ccv}\right] = 0, \qquad (24)$$

which is equivalent to the following eigen-energy equation:

$$\left(\varepsilon_k^c + E_{X\,q,q} - \omega\right)\delta_{kp} + F_{kqqp}^{cvvc} + w_{kqpq}^{cccc} = 0. \qquad (25)$$

We assume that exciton is created in one of two equivalent K-points, which correspond to the direct bandgap transition, therefore we put **q** equal to the K-point momentum. While, it is easy to generalize the solution on the case of arbitrary exciton momenta, the trion energy obtained with this approximation is sufficient to estimate the energy scale of the trionic effects in the system, including the position of the trion peak in the optical absorption spectrum.

The results of the solution of Eq. (25) in the case of contact kernel with A=0.395 (the case of the experimental exciton energy) at different values of the electron-electron screening are summarized in Table IV. As it follows from this Table, the results are very sensitive to the value of $\varepsilon$, though one can successfully reproduce the experimental result 20meV at a reasonable value $\varepsilon = 4.146$. We did not find a finite binding energy in the case of LDA, GEA and GGAs, a naturally expected result after extremely low excitonic energies for these kernels. Other two local kernels, LDA and GGA, do not give a finite binding energy either (see Table II, where all the results for the local kernels are summarized). The results in Tables I and II suggest that while the TDDFT exciton energies are mostly defined by local electron-hole attraction, in the case of trions in order to get a bound state of an electron and exciton one needs to take into account the long-range character of the interaction. Indeed, the electron-exciton interaction is more of "a dipole" type, contrary to the Coulomb interaction of the electron and hole. Similar to the exciton case, the fact that the contact kernel (2D LDA) results in a finite trion binding energy, contrary to the bulk LDA, suggests that the spatial constraint (charge non-homogeneity) of the system is important in this case too.

Table IV. The trion binding energy (in meV) in the case of local kernel with A=0.395 (corresponding to $E_X = 1000meV$) and different values of the electron-electron screening parameter. The corresponding exciton energy is 1,093meV.

| $\varepsilon$ | 2.8 | 3.3 | 4 | 4.146 | 4.2 | 4.5 | 5 | exp |
|---|---|---|---|---|---|---|---|---|
| $E_T$ | 0 | 0.15 | 1.74 | 20 | 31.7 | 105 | 223 | 20 |

Once again, in the case of LR kernels the results for the trion energy is rather sensitive to the value of the electron-electron screening. In particular, one can easily reproduce the experimental binding energy in the Slater case at fair value $\varepsilon = 2.7874$ (Table I).

Table VI. The trion binding energy (in meV) in the case of Slater kernel and different values of the electron-electron screening parameter. The corresponding exciton energy is 1,093meV.

| $\varepsilon$ | 1 | 2.7874 | 2.8 | 3.3 | 4.2 | exp |
|---|---|---|---|---|---|---|
| $E_T$ | 0 | 20 | 27 | 293 | 642 | 20 |

# IV. Conclusions

In this paper, we formulated a density-matrix TDDFT approach to study the exciton and trion effects in the bulk- and in nano-materials. This approach is physically transparent, with the inter-particle interaction defined by the TDDFT XC kernel. Similar to the exciton case, it has several advantages comparing to the standard many-body approaches – simplicity and accurately taken into account many-body correlation (especially screening) effects.

We applied the approach to study the exciton and trion binding energies in monolayer $MoS_2$. There are experimental indications that the corresponding binding energies are rather large (~1eV[8] and 0.02eV[7]), which makes it possible to use the exciton and trion effects at room temperatures. We have found a theoretical confirmation of these high binding energies in the case of a long-range Slater XC kernel, which takes into account correctly the nature of the electron-hole interaction as well as in the case of some phenomenological kernels: a LR kernel and contact kernel at physically reasonable values of the parameters. On the other hand, we found that, similar to excitons, one cannot obtain finite trion energies with standard LDA and GGA kernels due to missed LR nature of the electron-hole interaction in these cases.

The formalism described above can be used to study the binding energies and the ultrafast processes that involve excitonic, trionic and biexcitonic effects. The scheme proposed in the paper can be easily generalized on study of bound states with larger number of particles.

# Acknowledgements

This work was financially supported in part by US Department of Energy (DE-FG02-07ER46354). A.R.T. would like to acknowledge CONACYT (Mexico) for a support through the Postdoctoral Fellowship Program (number 184722).